\documentclass{article}
\usepackage{epsfig}
\newcommand{\mysection}{\setcounter{equation}{0}\section}

\def\beq{\begin{equation}}
\def\eeq{\end{equation}}
\def\beqa{\begin{eqnarray}}
\def\eeqa{\end{eqnarray}}

\newlength{\dinwidth} \newlength{\dinmargin}
\begin{document}
\begin{center}
{\Large \bf PRODUCTION OF LARGE TRANSVERSE MOMENTUM $W$ BOSONS AT THE 
TEVATRON}
\end{center}
\vspace{2mm}
\begin{center}
{\large Nikolaos Kidonakis$^1$ and Agust{\' \i}n Sabio Vera$^2$}\\
\vspace{2mm}
$^1$ {\it Kennesaw State University, 1000 Chastain Rd., \#1202 \\ 
Kennesaw, GA 30144--5591, USA}\\
$^2$ {\it II. Institut f{\" u}r Theoretische Physik, Universit{\"a}t Hamburg,\\
Luruper Chaussee 149, 22761 Hamburg, Germany}
\end{center}

\vspace{3mm}

\begin{center}
{\large ABSTRACT}
\end{center}

\vspace{3mm}

 We discuss the production of $W$ bosons at large transverse momentum
in $p{\bar p}$ collisions at the Tevatron Run I and II. 
$W$ boson hadroproduction is a process of 
relevance in testing the Standard Model and in estimates of backgrounds to new 
physics. The next--to--leading order 
cross section in the region of large transverse momentum 
is dominated by threshold soft--gluon corrections which arise from incomplete 
cancellations near partonic threshold between graphs with real emission 
and virtual graphs due to the limited phase space available for real 
gluon emission.
In this contribution it is shown how, by including 
next--to--next--to--leading--order (NNLO)
soft-gluon corrections, the transverse momentum distribution of the $W$ at the 
Tevatron is modestly enhanced and the dependence on the factorization 
and renormalization scales is significantly reduced. 

\vspace{5mm}

\mysection{INTRODUCTION}

At hadron--hadron colliders a process of major relevance is the production of 
$W$ bosons in the electroweak sector. Such a scattering process
 serves as a direct test of the Standard Model 
and as an estimate of backgrounds to possible new physical phenomena.
A significant example is 
$Wb{\bar b}$ production, the principal background to the associated 
Higgs boson production $p{\bar p}\rightarrow H(\rightarrow b{\bar b})W$ 
at the Tevatron~\cite{EV}. 

The full next--to--leading-order (NLO) calculation of the 
cross section for $W$ hadroproduction at large transverse momentum was 
first presented in Refs.~\cite{AR,gpw}.
The results in that work showed that the differential distributions 
in transverse momentum $Q_T$ of the $W$ boson are enhanced with respect 
to the leading order (LO) calculation. 
This $Q_T$ distribution falls rapidly with increasing $Q_T$, spanning five
orders of magnitude in the 30 GeV $< Q_T <$ 190 GeV region 
at the Tevatron. As expected from a consistent perturbative expansion, the
 NLO corrections also considerably stabilize
the dependence of the cross section on the factorization and renormalization
scales.

When a hard--scattering cross section takes place close to the partonic 
threshold, as is the case in high transverse momentum production of 
electroweak bosons, corrections related to the emission of soft gluons from 
the partons in the process have to be taken into account. Near partonic 
threshold there are incomplete cancellations between real emission graphs 
and virtual ones, generating, at each order in perturbation theory, large 
logarithms stemming from the limited phase space available for real gluon 
emission near threshold. When these threshold corrections are calculated
in the eikonal approximation they exponentiate as a consequence of the 
factorization properties~\cite{KS,KOS,LOS} of the cross section. In general 
terms the factorization of the cross section is done in terms of  
functions which describe those gluons collinear to the incoming partons, 
hard quanta, and noncollinear soft--gluons. The renormalization
group properties of such functions result in resummation.
The threshold corrections have by now been successfully resummed for many
processes~\cite{NK}.

In this chapter the resummation of threshold logarithms 
together with the expansion of the resummed cross section at
next--to--next--to--leading order (NNLO)
for electroweak $W$ boson hadroproduction, which was first
presented in Ref.~\cite{NKVD} at next--to--next--to--leading logarithmic 
(NNLL) accuracy, are further studied.  We note that in Ref.~\cite{NKVD} 
no numerical phenomenological studies were made.

Quite recently, an approach which unifies the calculation of NNLO soft and
virtual corrections for hadron--hadron
and lepton--hadron processes has been presented in Ref.~\cite{NKuni}.
This work puts together and extends previous approaches by going beyond
NNLL accuracy.
In this chapter that work is closely followed to calculate the NNLO soft 
corrections
to $W$ production at large transverse momentum at the Tevatron.
The theoretical results presented here are consistent with those in 
Ref.~\cite{NKVD} although they slightly differ in that  
the transverse momentum $Q_T$ is now used as the hard scale instead 
of the mass of the $W$.

This chapter is based on the work of Ref.~\cite{NKASV} where 
we studied the significance of NNLO soft--gluon corrections
for $W$--boson production at large transverse momentum. 
The accuracy of the theoretical calculation is increased to
next--to--next--to--next--to--leading logarithms (NNNLL). 
We note that we work in the $\overline{\rm MS}$ scheme throughout.

The present results are closely related to previous studies of direct photon 
production, for which numerical analysis 
were presented in Ref.~\cite{NKJO}.
The relation arises because the partonic subprocesses involved in direct 
photon production are similar to the ones discussed in this chapter.

In Section 2 the kinematics of the partonic
subprocesses involved is discussed and the corrections to be calculated 
are defined.
In Section 3 results for the NLO soft and virtual, and 
the  NNLO soft--gluon corrections are provided. It is shown how the NLO soft 
corrections agree near partonic threshold with the 
exact NLO calculations of Refs.~\cite{AR,gpw} while our 
NNLO soft--gluon corrections provide new predictions.
In Section 4 we study the numerical effect of the NNLO soft--gluon
corrections to $W$ hadroproduction at the Tevatron Run I and II,
while also showing that the NLO cross section is dominated by soft--gluon 
corrections.

\mysection{KINEMATICS FOR $W$ PRODUCTION IN HADRON COLLISIONS}

We study the production of a $W$ boson, with mass $m_W$,
in collisions of hadron $h_A$ with hadron $h_B$,
\beq
h_A(P_A)+h_B(P_B) \longrightarrow W(Q) + X \, ,
\eeq
where $X$ denotes any allowed additional final-state particles.
At the parton level, the lowest--order subprocesses for the 
production of a $W$ boson and a parton are 
\beqa
q(p_a)+g(p_b) &\longrightarrow& W(Q) + q(p_c)  \, ,
\nonumber \\
q(p_a)+{\bar q}(p_b) &\longrightarrow& W(Q) + g(p_c)  \, .
\label{partsub}
\eeqa

We can write the factorized single--particle--inclusive cross section as
\beqa
E_Q\,\frac{d\sigma_{h_Ah_B{\rightarrow}W(Q)+X}}{d^3Q} &=&
\sum_{f} \int dx_1 \, dx_2 \; \phi_{f_a/h_A}(x_1,\mu_F^2) 
\; \phi_{f_b/h_B}(x_2,\mu_F^2) 
\nonumber \\ && \hspace{-10mm} \times \,
E_Q\,\frac{d\hat{\sigma}_{f_af_b{\rightarrow}W(Q)+X}}{d^3Q}
(s,t,u,Q,\mu_F,\alpha_s(\mu_R^2)) \label{factor}
\label{factW}
\eeqa
where $E_Q=Q^0$, $\phi_{f/h}(x)$ is the parton distribution for parton 
$f$ carrying a momentum fraction $x$ in hadron $h$,
and $\hat{\sigma}$ is the perturbative parton--level cross section. 
The initial--state collinear singularities are factorized into the
parton distributions at factorization scale $\mu_F$, while $\mu_R$ is the
renormalization scale.

The hadronic and partonic kinematical invariants in the process are 
\beqa
S &=& (P_A+P_B)^2, \nonumber\\
T &=& (P_A-Q)^2, \nonumber\\
U &=& (P_B-Q)^2, \nonumber\\
S_2 &=& S + T + U - Q^2, \nonumber\\
s &=& (p_a+p_b)^2, \nonumber\\
t &=& (p_a-Q)^2, \nonumber\\ 
u &=& (p_b-Q)^2, \nonumber\\
s_2 &=& s + t + u - Q^2,
\label{partkin} 
\eeqa
where $S_2$ and $s_2$ are the invariant masses of the system recoiling 
against the electroweak boson at the hadron and parton levels, respectively.
The invariant $s_2=(p_a+p_b-Q)^2$ parametrizes the inelasticity of the 
parton scattering, taking the value $s_2=0$ at partonic threshold.
Since $x_i$ is the initial--parton momentum fraction, defined
by $p_a = x_1 P_A$ and $p_b = x_2 P_B$, the hadronic and partonic kinematical 
invariants are related by 
\beqa
s &=& x_1 x_2 S, \nonumber\\
t-Q^2 &=& x_1 (T-Q^2) \nonumber\\
u-Q^2 &=& x_2 (U-Q^2). 
\eeqa

We note that for numerical calculations it is convenient to 
write the hadronic kinematical variables $T$ and $U$ in the alternative way
\begin{eqnarray}
T &=& m_W^2 - m_T \sqrt{S} e^{-y} \nonumber\\
U &=& m_W^2 - m_T \sqrt{S} e^{y},
\end{eqnarray}
where $m_W$ is the mass of the $W$ boson, $m_T = \sqrt{Q_T^2 + m_W^2}$ is 
its transverse mass,
and $y$ is its rapidity. With this notation the differential $Q_T$
distribution can then be expressed as
\begin{eqnarray} \frac{d \sigma_{h_A h_B\rightarrow W+X}}{d Q_T^2}
\left(S,m_W^2,Q_T\right) &=& \sum_f \int_0^1 dy' \int^1_A dx_1
\int^{s_2^{\rm max}}_0 d s_2 \frac{2 \pi Y}{x_1 S-\sqrt{S} m_T e^y} \\
&&\hspace{-1cm} \times \phi_{f_a/h_A}\left(x_1,\mu_F^2\right)
\phi_{f_b/h_B}\left(x_2,\mu_F^2\right)
E_Q \frac{d{\hat \sigma}_{f_a f_b\rightarrow W+ X}}{d^3 Q}
\left(x_1,x_2,y\right) \nonumber,
\end{eqnarray}
where the relations $Y = \ln{\left(B+\sqrt{B^2-1}\right)}$, 
$B = (S+m_W^2)/(2 m_T \sqrt{S})$, and 
$y= Y (2 y'-1)$, hold. The kinematical limits of the integrations read
\beqa
s_2^{\rm max} &=& m_W^2-\sqrt{S}m_T e^y + x_1
\left(S-\sqrt{S}m_T e^{-y}\right),\\
A &=& \frac{\sqrt{S}m_T e^y -m_W^2}{S-\sqrt{S} m_T e^{-y}},\quad
x_2 = \frac{s_2-m_W^2+\sqrt{S}m_T x_1 e^{-y}}{x_1 S - \sqrt{S} m_T e^y}.
\eeqa

In general, the partonic cross section $\hat{\sigma}$ 
includes distributions with respect 
to $s_2$ at $n$--th order in the strong coupling $\alpha_s$ of the type
\beq
\left[\frac{\ln^{m}(s_2/Q_T^2)}{s_2} \right]_+, \hspace{10mm} m\le 2n-1\, ,
\label{zplus}
\eeq
defined by their integral with any smooth function $f$ by 
\beqa
\int_0^{s_{2 \, max}} ds_2 \, f(s_2) \left[\frac{\ln^m(s_2/Q_T^2)}
{s_2}\right]_{+} &\equiv& \nonumber\\
&&\hspace{-6cm}
\int_0^{s_{2\, max}} ds_2 \frac{\ln^m(s_2/Q_T^2)}{s_2} [f(s_2) - f(0)] 
+\frac{1}{m+1} \ln^{m+1}\left(\frac{s_{2\, max}}{Q_T^2}\right) f(0) \, .
\label{splus}
\eeqa
These ``plus'' distributions are the remnants of cancellations between
real and virtual contributions to the cross section. 
Note that in Ref. \cite{NKVD} $Q$ was used instead of $Q_T$ in the
plus distributions. Here we prefer to use $Q_T$ as we find it a slightly
better physical hard scale for the $Q_T$ distributions that we will
be calculating.    
Below we will make use of the terminology that at $n$--th order in $\alpha_s$
the leading logarithms (LL) 
are those with $m=2n-1$ in Eq. (\ref{zplus}), 
next--to--leading logarithms (NLL)
with $m=2n-2$, next--to--next--to--leading logarithms (NNLL) with $m=2n-3$,
and next--to--next--to--next--to--leading logarithms (NNNLL) with $m=2n-4$.

\mysection{NEXT--TO--NEXT--TO--LEADING ORDER SOFT--GLUON CORRECTIONS}

In this section, we first present the NLO soft and virtual corrections
for each of the subprocesses in Eq. (\ref{partsub}).
We then present the NNLO soft and some virtual corrections. 

\subsection{NLO and NNLO corrections for $qg \longrightarrow Wq$}

We begin with the $qg \longrightarrow Wq$ subprocess.
The Born differential cross section for this process is
\beq
E_Q \frac{d\sigma^B_{qg\rightarrow Wq}}{d^3Q}
=F^B_{qg \rightarrow Wq} \, \delta(s_2) \, ,
\eeq
where
\beqa
F^B_{qg \rightarrow Wq} &=& \frac{\alpha \, 
\alpha_s(\mu_R^2)C_F}{s(N_c^2-1)}
A^{qg} \, \sum_{f} |L_{ff_a}|^2 \, ,\\
A^{qg} &=& - \left(\frac{s}{t}+\frac{t}{s}+\frac{2uQ^2}{st}\right) \, , 
\nonumber
\eeqa
with $L$ the left--handed couplings of the
$W$ boson to the quark line, $f$ the quark flavor and $\sum_f$
the sum over the flavors allowed by the CKM mixing and by the energy
threshold. For the $L$ couplings we choose the conventions of
Ref.~\cite{gpw}. Also $C_F=(N_c^2-1)/(2N_c)$ with $N_c=3$ the number
of colors.

We can write the NLO soft and virtual corrections for 
$qg \longrightarrow Wq$ 
in single--particle inclusive kinematics as
\beq
E_Q\frac{d{\hat\sigma}^{(1)}_{qg \rightarrow Wq}}{d^3Q} = 
F^B_{qg \rightarrow Wq} 
{\alpha_s(\mu_R^2)\over\pi}\,
\left\{c_3^{qg} \, \left[\frac{\ln(s_2/Q_T^2)}{s_2}\right]_+ 
+c_2^{qg} \, \left[\frac{1}{s_2}\right]_+ + c_1^{qg} \, 
\delta(s_2)\right\} \, .
\label{qgnlo}
\eeq

Note that the $[\ln(s_2/Q_T^2)/s_2]_+$ term (which is the LL, since
$m=n=1$ in Eq.~(\ref{zplus}))
and the $[1/s_2]_+$ term (NLL, since $n=1$, $m=0$) 
are the soft gluon corrections.
The $\delta(s_2)$ term is the contribution from the virtual corrections.
Below we use the terminology ``NLO--NLL'' to indicate when,
at next--to--leading order, we include the LL and NLL soft--gluon terms 
(as well as scale-dependent terms in $\delta(s_2)$). 
Also the terminology ``soft and virtual'' is used to denote
all the terms in Eq. (\ref{qgnlo}).
As we will see, we need to know both soft and virtual corrections at NLO
in order to derive the soft terms at NNLO with at least NNLL accuracy.

The NLO coefficients in Eq. (\ref{qgnlo}) are 
$c_3^{qg}=C_F+2C_A$,
\beq
c_2^{qg}=-\left(C_F + C_A\right) \ln\left(\frac{\mu_F^2}{Q_T^2}\right)
- {3\over 4} C_F - C_A \ln{\left(t u \over s Q_T^2\right)} \, ,
\eeq
and
\beq
c_1^{qg}=\frac{1}{2A^{qg}}\left[B_1^{qg}+B_2^{qg} n_f
+C_1^{qg}+C_2^{qg} n_f \right]+\frac{c_3^{qg}}{2}
\ln^2\left(\frac{Q_T^2}{Q^2}\right)
+c_2^{qg} \ln\left(\frac{Q_T^2}{Q^2}\right)\, , 
\eeq
with $C_A=N_c$, $n_f=5$ the number of light quark flavors, and
$B_1^{qg}$, $B_2^{qg}$, $C_1^{qg}$, and $C_2^{qg}$ 
as given in the Appendix
of Ref. \cite{gpw} but without the renormalization counterterms
and using $f_A \equiv\ln(A/Q^2)=0$. 

Note that we can write $c_2^{qg} \equiv c_{2 \; \mu}^{qg}+T_2^{qg}$
with $c_{2 \; \mu}^{qg} \equiv -(C_F+C_A)\ln(\mu_F^2/s)$.
Similarly we also write $c_1^{qg}\equiv c_{1 \; \mu}^{qg}+T_1^{qg}$ with
\beq
c_{1\; \mu}^{qg} \equiv \ln\left(\frac{\mu_F^2}{s}\right)
\left[-\frac{\beta_0}{4}+C_F\left(\ln\left(\frac{-u}{Q_T^2}\right)
-\frac{3}{4}\right)+C_A\ln\left(\frac{-t}{Q_T^2}\right)\right] 
+\frac{\beta_0}{4} \ln\left(\frac{\mu_R^2}{s}\right) \, ,
\eeq
where $\beta_0=(11C_A-2n_f)/3$ is the lowest--order beta function.
Thus,
$c_{2 \; \mu}^{qg}$ and $c_{1 \; \mu}^{qg}$ are scale--dependent parts of the
$c_2^{qg}$ and $c_1^{qg}$ coefficients, respectively, while
$T_2^{qg}$ and $T_1^{qg}$ are scale--independent parts.  
We have kept the factorization scale $\mu_F$ and the renormalization scale
$\mu_R$ separate.
Finally, another useful notation is
$c_1^{qg}\equiv {c'}_{1 \; \mu}^{qg}+{T'}_1^{qg}$
where ${c'}_{1 \; \mu}^{qg}$ is defined as $c_{1\; \mu}^{qg}$
with $Q_T^2$ instead of $s$ in the denominators of the logarithms
involving the scales $\mu_F$ and $\mu_R$.

Using the above conventions, the NNLO soft corrections for
$qg \longrightarrow Wq$ can be written as
\beq
E_Q\frac{d{\hat\sigma}^{(2)}_{qg \rightarrow Wq}}{d^3Q} = 
F^B_{qg \rightarrow Wq} 
\frac{\alpha_s^2(\mu_R^2)}{\pi^2} \, 
{\hat{\sigma'}}^{(2)}_{qg \rightarrow Wq}
\label{NNLOmqg}
\eeq
with
\beqa
{\hat{\sigma'}}^{(2)}_{qg \rightarrow Wq}&=& 
\frac{1}{2} (c_3^{qg})^2 \, \left[\frac{\ln^3(s_2/Q_T^2)}{s_2}\right]_+ 
+\left[\frac{3}{2} c_3^{qg} \, c_2^{qg} 
- \frac{\beta_0}{4} c_3^{qg}
+C_F \frac{\beta_0}{8}\right] \left[\frac{\ln^2(s_2/Q_T^2)}{s_2}\right]_+
\nonumber \\ && \hspace{-25mm}
{}+\left\{c_3^{qg} \, c_1^{qg} +(c_2^{qg})^2
-\zeta_2 \, (c_3^{qg})^2 -\frac{\beta_0}{2} \, T_2^{qg} 
+\frac{\beta_0}{4} c_3^{qg}  \ln\left(\frac{\mu_R^2}{s}\right)
+(C_F+C_A) \, K \right.
\nonumber \\ && \hspace{-25mm} \quad \quad \left.
{}+C_F \left[-\frac{K}{2} 
+\frac{\beta_0}{4} \, \ln\left(\frac{Q_T^2}{s}\right)\right]
-\frac{3}{16} \beta_0 C_F \right\}
\left[\frac{\ln(s_2/Q_T^2)}{s_2}\right]_+
\nonumber \\ && \hspace{-25mm} 
{}+\left\{c_2^{qg} \, c_1^{qg} -\zeta_2 \, c_2^{qg} \, c_3^{qg}
+\zeta_3 \, (c_3^{qg})^2 
-\frac{\beta_0}{2} T_1^{qg}
+\frac{\beta_0}{4}\, c_2^{qg} \ln\left(\frac{\mu_R^2}{s}\right) 
+{\cal G}_{qg}^{(2)}
\right. 
\nonumber \\ && \hspace{-25mm} \quad \quad
{}+(C_F+C_A) \left[\frac{\beta_0}{8} 
\ln^2\left(\frac{\mu_F^2}{s}\right)
-\frac{K}{2}\ln\left(\frac{\mu_F^2}{s}\right)\right]
-C_F\, K \, \ln\left(\frac{-u}{Q_T^2}\right)
-C_A\, K \, \ln\left(\frac{-t}{Q_T^2}\right)
\nonumber \\ && \hspace{-25mm} \quad \quad \left.
{}+C_F \, \left[\frac{\beta_0}{8}
\ln^2\left(\frac{Q_T^2}{s}\right)
-\frac{K}{2}\ln\left(\frac{Q_T^2}{s}\right)\right]
-\frac{3}{16}\beta_0 C_F\, \ln\left(\frac{Q_T^2}{s}\right) \right\}
\left[\frac{1}{s_2}\right]_+  \, ,
\eeqa
where $K=C_A(67/18-\zeta_2)-5n_f/9$ is a two--loop function
in the $\overline{\rm MS}$ scheme \cite{KodTr}, $\zeta_2=\pi^2/6$ and
$\zeta_3=1.2020569\cdots$ are Riemann zeta functions, 
$\beta_1=34 C_A^2/3 \, - \, 2n_f(C_F+5C_A/3)$ is the 
next--to--leading order beta function, and
\beq
{\gamma'}_{q/q}^{(2)}=C_F^2\left(\frac{3}{32}-\frac{3}{4}\zeta_2
+\frac{3}{2}\zeta_3\right)
+C_F C_A\left(-\frac{3}{4}\zeta_3+\frac{11}{12}\zeta_2+\frac{17}{96}\right)
+n_f C_F \left(-\frac{\zeta_2}{6}-\frac{1}{48}\right)\, ,
\eeq
\beq
{\gamma'}_{g/g}^{(2)}=C_A^2\left(\frac{2}{3}+\frac{3}{4}\zeta_3\right)
-n_f\left(\frac{C_F}{8}+\frac{C_A}{6}\right) \, , 
\eeq
are two--loop parton anomalous dimensions \cite{GALY,GFP}.

The LL are the  $[\ln^3(s_2/Q_T^2)/s_2]_+$ term (since $n=2$, $m=3$ 
in Eq.~(\ref{zplus})), the NLL are the 
$[\ln^2(s_2/Q_T^2)/s_2]_+$ term ($n=2$, $m=2$), the NNLL are the 
$[\ln(s_2/Q_T^2)/s_2]_+$ term ($n=2$, $m=1$), and the NNNLL are the
$[1/s_2]_+$ term ($n=2$, $m=0$).
The function ${\cal G}^{(2)}_{q g}$ in the NNNLL term denotes a set of 
two--loop contributions \cite{NK2l} and is given by
\cite{NKuni,NKASV}
\beqa
{\cal G}^{(2)}_{q g}&=&C_F^2\left(-\frac{3}{32}+\frac{3}{4}\zeta_2
-\frac{3}{2}\zeta_3\right)+ C_F C_A \left(\frac{3}{4} \zeta_3
-\frac{11}{12}\zeta_2-\frac{189}{32}\right)
\nonumber \\ && \hspace{-15mm}
{}+C_A^2 \left(\frac{7}{4} \zeta_3
+\frac{11}{3}\zeta_2-\frac{41}{216}\right)
+ n_f C_F \left(\frac{1}{6}\zeta_2+\frac{17}{16}\right) 
+n_f C_A \left(-\frac{2}{3}\zeta_2
-\frac{5}{108}\right) \, .
\eeqa
Note that we have not included in ${\cal G}^{(2)}_{q g}$ two--loop
process--dependent contributions. These additional contributions
have to be included for a full NNNLL calculation. 
However from related studies for
other processes, including top hadroproduction \cite{NKtop} 
and direct photon production \cite{NKJO}
we expect such contributions to be small. It is actually the 
$-\zeta_2 c_2^{qg} c_3^{qg}+\zeta_3 (c_3^{qg})^2$ terms
in Eq.~(\ref{NNLOqg}) that provide the major contribution to the NNNLL term.

We use the terminology ``NNLO--NNLL'' below, to indicate that we include
the LL, NLL, and NNLL terms at NNLO, while we use ``NNLO--NNNLL'' to
indicate that in addition we include the NNNLL terms as well. 

We remind the reader that here we do not calculate the full NNLO virtual 
corrections, proportional to $\delta(s_2)$. 
However, we have calculated explicitly here all the 
$\delta(s_2)$ terms that  include the factorization and renormalization 
scale--dependence. Note that when discussing scale dependence 
at NNLO--NNNLL we include all scale--dependent $\delta(s_2)$ terms,
which is a consistent approach from the
resummation procedure \cite{NKtop}. 
These terms are
\beqa
&& \hspace{-8mm}
{}\left\{\frac{1}{2}({c'}_{1 \; \mu}^{qg})^2
+{c'}_{1 \; \mu}^{qg} {T'}_{1}^{qg} 
+\frac{\beta_0}{4} {c'}_{1 \; \mu}^{qg} \ln\left(\frac{Q_T^2}{s}\right)
+\frac{\beta_0}{4} c_1^{qg} \ln\left(\frac{\mu_R^2}{Q_T^2}\right) 
-(C_F+C_A)^2 \frac{\zeta_2}{2} \ln^2\left(\frac{\mu_F^2}{Q_T^2}\right)
\right.
\nonumber \\ && \hspace{-6mm}
{}+(C_F+C_A) \ln\left(\frac{\mu_F^2}{Q_T^2}\right)
\left(\zeta_2 T_2^{qg}-\zeta_2 (C_F+C_A)\ln\left(\frac{Q_T^2}{s}\right)
-\zeta_3 c_3^{qg}\right)
\nonumber \\ && \hspace{-6mm} 
{}-\frac{\beta_0^2}{32} \ln^2\left(\frac{\mu_R^2}{Q_T^2}\right)
-\frac{\beta_0^2}{16} \ln\left(\frac{\mu_R^2}{Q_T^2}\right)
\ln\left(\frac{Q_T^2}{s}\right)
+\frac{\beta_1}{16}\ln\left(\frac{\mu_R^2}{Q_T^2}\right)
\nonumber \\ && \hspace{-6mm}  
{}+\frac{\beta_0}{8} \left[\frac{3}{4}C_F+\frac{\beta_0}{4}
-C_F \, \ln\left(\frac{-u}{Q_T^2}\right)
-C_A \, \ln\left(\frac{-t}{Q_T^2}\right)\right] 
\left[\ln^2\left(\frac{\mu_F^2}{Q_T^2}\right)
+2\ln\left(\frac{\mu_F^2}{Q_T^2}\right)\ln\left(\frac{Q_T^2}{s}\right)\right]
\nonumber \\ && \hspace{-6mm}  
{}+C_F \frac{K}{2} \, \ln\left(\frac{-u}{Q_T^2}\right)
\ln\left(\frac{\mu_F^2}{Q_T^2}\right)
+C_A \frac{K}{2} \, \ln\left(\frac{-t}{Q_T^2}\right)
\ln\left(\frac{\mu_F^2}{Q_T^2}\right)
\nonumber \\ && \hspace{-6mm}  \left.
{}-({\gamma'}_{q/q}^{(2)}+{\gamma'}_{g/g}^{(2)}) 
\ln\left(\frac{\mu_F^2}{Q_T^2}\right)\right\} \delta(s_2) \, .
\label{NNLOqg}
\eeqa

\subsection{NLO and NNLO corrections for $q{\bar q} \longrightarrow Wg$}

Next, we study  the partonic subprocess $q{\bar q} \longrightarrow Wg$.
The Born differential cross section is
\beq
E_Q \frac{d\sigma^B_{q {\bar q}\rightarrow Wg}}{d^3Q}
=F^B_{q{\bar q} \rightarrow Wg} \, \delta(s_2) \, ,
\eeq
where
\beqa
F^B_{q{\bar q} \rightarrow Wg} &=&\frac{\alpha \alpha_s(\mu_R^2)C_F}{sN_c}
A^{q\bar q}\,  |L_{f_bf_a}|^2 \, , \\
A^{q\bar q} &=& \frac{u}{t}+\frac{t}{u}+\frac{2Q^2s}{tu} \, .
\nonumber
\eeqa 

The NLO soft and virtual corrections in single--particle inclusive kinematics
can be written as 
\begin{equation}
E_Q\frac{d{\hat\sigma}^{(1)}_{q{\bar q} \rightarrow Wg}}{d^3Q} =
{F^B_{q{\bar q} \rightarrow Wg}} {\alpha_s(\mu_R^2)\over\pi}\,
\left\{c_3^{q \bar q} \, \left[\frac{\ln(s_2/Q_T^2)}{s_2}\right]_+ 
+c_2^{q \bar q} \, \left[\frac{1}{s_2}\right]_+ 
+c_1^{q \bar q} \, \delta(s_2)\right\} \, .
\label{qqbarnlo}
\end{equation}
Here the NLO coefficients are
$c_3^{q \bar q}=4C_F-C_A$,
\beq
c_2^{q \bar q}=- 2 C_F \ln\left(\frac{\mu_F^2}{Q_T^2}\right) 
- \left(2 C_F- C_A \right) \ln\left(\frac{t u}{s Q_T^2}\right) 
-\frac{\beta_0}{4}
\eeq
and
\beq
c_1^{q \bar q}=\frac{1}{2A^{q \bar q}}\left[B_1^{q \bar q}+C_1^{q \bar q}
+(B_2^{q \bar q}+D_{aa}^{(0)}) \, n_f \right] 
+\frac{c_3^{q \bar q}}{2}
\ln^2\left(\frac{Q_T^2}{Q^2}\right)
+c_2^{q \bar q} \ln\left(\frac{Q_T^2}{Q^2}\right)\, ,
\eeq
with $B_1^{q \bar q}$, $B_2^{q \bar q}$,
$C_1^{q \bar q}$, and $D_{aa}^{(0)}$
as given in the Appendix of Ref. \cite{gpw} but without the renormalization 
counterterms and using $f_A=0$.
Again, we can write $c_1^{q \bar q}=c_{1 \; \mu}^{q \bar q}+T_1^{q \bar q}$
with
\beq
c_{1 \; \mu}^{q \bar q}=\ln\left(\frac{\mu_F^2}{s}\right)  
C_F\left[\ln\left(\frac{tu}{Q_T^4}\right)-\frac{3}{2}\right]
+\frac{\beta_0}{4} \ln\left(\frac{\mu_R^2}{s}\right) \, ,
\eeq
and $c_2^{q \bar q}=c_{2 \; \mu}^{q \bar q}+T_2^{q \bar q}$ 
with $c_{2 \; \mu}^{q \bar q}=- 2 C_F \ln(\mu_F^2/s)$. 
Finally, another useful notation is
$c_1^{q \bar q}\equiv {c'}_{1 \; \mu}^{q \bar q}+{T'}_1^{q \bar q}$
where ${c'}_{1 \; \mu}^{q \bar q}$ is defined as $c_{1\; \mu}^{q \bar q}$
with $Q_T^2$ instead of $s$ in the denominators of the logarithms
involving the scales $\mu_F$ and $\mu_R$.

The NNLO soft corrections 
for $q{\bar q} \longrightarrow Wg$ can be written as
\beq
E_Q\frac{d{\hat\sigma}^{(2)}_{q{\bar q} \rightarrow Wg}}{d^3Q} =
{F^B_{q{\bar q} \rightarrow Wg}}
\frac{\alpha_s^2(\mu_R^2)}{\pi^2} \, {\hat{\sigma'}}^{(2)}_{q{\bar q} 
\rightarrow Wg}
\label{NNLOmqqb}
\eeq
with
\beqa
{\hat{\sigma'}}^{(2)}_{q{\bar q} \rightarrow Wg}&=& 
\frac{1}{2} (c_3^{q \bar q})^2 \, \left[\frac{\ln^3(s_2/Q_T^2)}{s_2}\right]_+ 
+\left[\frac{3}{2} c_3^{q \bar q} \, c_2^{q \bar q} 
- \frac{\beta_0}{4} c_3^{q \bar q}
+C_A \frac{\beta_0}{8}\right] \left[\frac{\ln^2(s_2/Q_T^2)}{s_2}\right]_+ 
\nonumber \\ && \hspace{-15mm}
{}+\left\{c_3^{q \bar q} \, c_1^{q \bar q} +(c_2^{q \bar q})^2
-\zeta_2 \, (c_3^{q \bar q})^2 -\frac{\beta_0}{2} \, T_2^{q \bar q} 
+\frac{\beta_0}{4} c_3^{q \bar q}  \ln\left(\frac{\mu_R^2}{s}\right)
+2C_F \, K \right.
\nonumber \\ && \hspace{-15mm} \quad \quad \left.
{}+C_A \left[-\frac{K}{2} 
+\frac{\beta_0}{4} \, \ln\left(\frac{Q_T^2}{s}\right)\right]
-\frac{\beta_0^2}{16} \right\}
\left[\frac{\ln(s_2/Q_T^2)}{s_2}\right]_+ 
\nonumber \\ && \hspace{-15mm} 
{}+\left\{c_2^{q \bar q} \, c_1^{q \bar q} -\zeta_2 \, c_2^{q \bar q}
 \, c_3^{q \bar q}+\zeta_3 \, (c_3^{q \bar q})^2 
-\frac{\beta_0}{2} T_1^{q \bar q}
+\frac{\beta_0}{4}\, c_2^{q \bar q} \ln\left(\frac{\mu_R^2}{s}\right) 
+{\cal G}^{(2)}_{q \bar q} \right. 
\nonumber \\ && \hspace{-15mm} \quad \quad
{}+C_F \left[\frac{\beta_0}{4} 
\ln^2\left(\frac{\mu_F^2}{s}\right)
-K\ln\left(\frac{\mu_F^2}{s}\right)
-K \, \ln\left(\frac{tu}{Q_T^4}\right)\right]
\nonumber \\ && \hspace{-15mm} \quad \quad \left.
{}+C_A \, \left[\frac{\beta_0}{8}
\ln^2\left(\frac{Q_T^2}{s}\right)
-\frac{K}{2}\ln\left(\frac{Q_T^2}{s}\right)\right]
-\frac{\beta_0^2}{16}\, \ln\left(\frac{Q_T^2}{s}\right) \right\}
\left[\frac{1}{s_2}\right]_+  \, .
\eeqa
The function ${\cal G}^{(2)}_{q {\bar q}}$ denotes again
a set of two--loop contributions \cite{NKuni,NKASV} and is given by
\beq
{\cal G}^{(2)}_{q {\bar q}}
=C_F C_A \left(\frac{7}{2} \zeta_3
+\frac{22}{3}\zeta_2-\frac{299}{27}\right)
+n_f C_F \left(-\frac{4}{3}\zeta_2+\frac{50}{27}\right) \, . 
\eeq
Again, we have not included in ${\cal G}^{(2)}_{q {\bar q}}$
all two--loop process--dependent contributions.

We also note that we do not calculate the full virtual corrections.
The NNLO $\delta(s_2)$ terms that we calculated only include the 
factorization and renormalization scale dependence
and are used in the study of the scale dependence of the NNLO--NNNLL
cross section. These terms are:
\beqa
&& \hspace{-16mm} \left\{\frac{1}{2}({c'}_{1 \; \mu}^{q \bar q})^2
+{c'}_{1 \; \mu}^{q \bar q} {T'}_{1}^{q \bar q}
+\frac{\beta_0}{4}{c'}_{1 \; \mu}^{q \bar q}
\ln\left(\frac{Q_T^2}{s}\right)  
+\frac{\beta_0}{4} c_1^{q \bar q} \ln\left(\frac{\mu_R^2}{Q_T^2}\right) 
\right.
\nonumber \\ && \hspace{-15mm} \quad \quad
{}-2C_F^2 \zeta_2 \ln^2\left(\frac{\mu_F^2}{Q_T^2}\right)
+2C_F \ln\left(\frac{\mu_F^2}{Q_T^2}\right)
\left(\zeta_2 T_2^{q \bar q}-2 \zeta_2 C_F\ln\left(\frac{Q_T^2}{s}\right)
-\zeta_3 c_3^{q \bar q}\right)
\nonumber \\ && \hspace{-15mm} \quad \quad
{}-\frac{\beta_0^2}{32} \ln^2\left(\frac{\mu_R^2}{Q_T^2}\right)
-\frac{\beta_0^2}{16} \ln\left(\frac{\mu_R^2}{Q_T^2}\right)
\ln\left(\frac{Q_T^2}{s}\right)
+\frac{\beta_1}{16}\ln\left(\frac{\mu_R^2}{Q_T^2}\right)
\nonumber \\ && \hspace{-15mm} \quad \quad 
{}+\frac{\beta_0}{8} \left[\frac{3}{2}C_F
-C_F \, \ln\left(\frac{tu}{Q_T^4}\right)\right] 
\left[\ln^2\left(\frac{\mu_F^2}{Q_T^2}\right)
+2\ln\left(\frac{\mu_F^2}{Q_T^2}\right)\ln\left(\frac{Q_T^2}{s}\right)\right]
\nonumber \\ && \hspace{-15mm} \left. \quad \quad 
{}+C_F \frac{K}{2} \, \ln\left(\frac{tu}{Q_T^4}\right)
\ln\left(\frac{\mu_F^2}{Q_T^2}\right)
-2{\gamma'}_{q/q}^{(2)} \ln\left(\frac{\mu_F^2}{Q_T^2}\right)
\right\} \delta(s_2) \, .
\label{NNLOqqb}
\eeqa

\mysection{NUMERICAL RESULTS FOR LARGE-$Q_T$ \\
$W$ PRODUCTION AT THE TEVATRON }

\begin{figure}[htb] 
\setlength{\epsfxsize=0.75\textwidth}
\setlength{\epsfysize=0.4\textheight}
\centerline{\epsffile{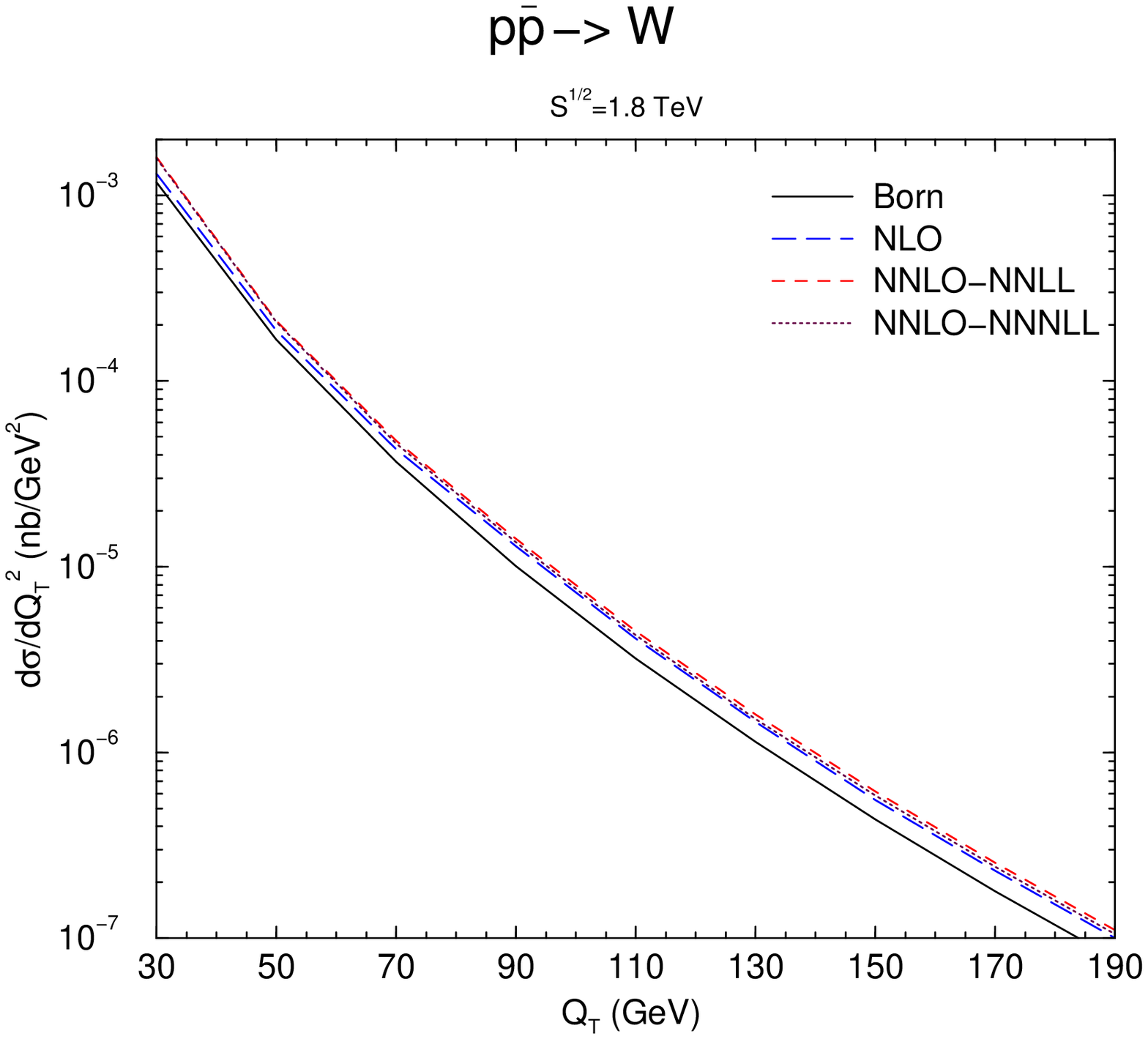}}
\caption[]{The differential cross section,
$d\sigma/dQ_T^2$, for $W$ hadroproduction in $p \bar p$ collisions
at the Tevatron with $\sqrt{S}=1.8$ TeV and $\mu_F=\mu_R=Q_T$.
Shown are the Born (solid line), NLO (long-dashed line),
NNLO--NNLL (short--dashed line), and NNLO--NNNLL (dotted line) results.
}
\label{fig1} 
\end{figure}

\begin{figure}[htb] 
\setlength{\epsfxsize=0.75\textwidth}
\setlength{\epsfysize=0.4\textheight}
\centerline{\epsffile{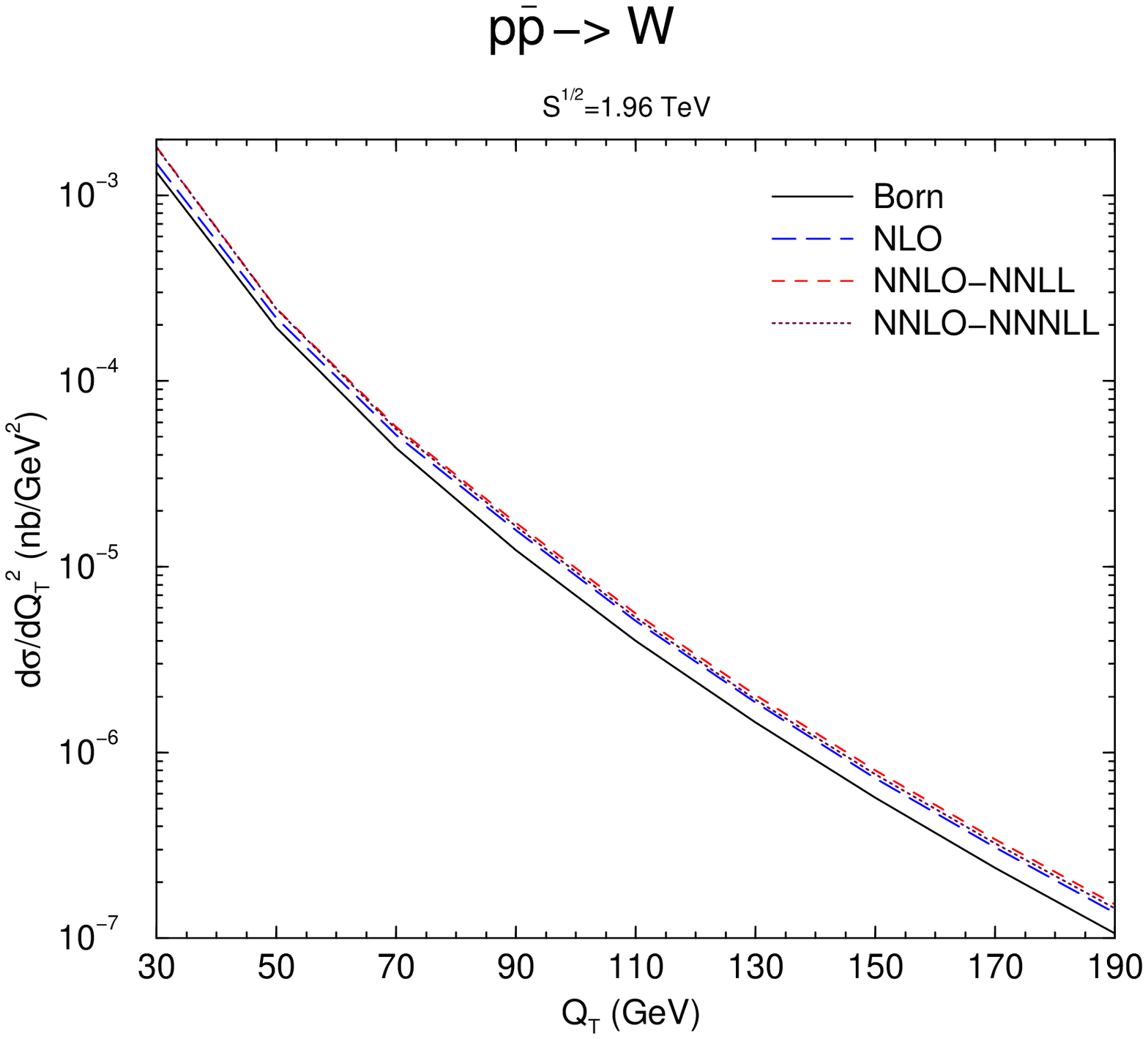}}
\caption[]{The differential cross section,
$d\sigma/dQ_T^2$, for $W$ hadroproduction in $p \bar p$ collisions
at the Tevatron Run II with $\sqrt{S}=1.96$ TeV and $\mu_F=\mu_R=Q_T$.
Shown are the Born (solid line), exact NLO (long--dashed line), 
NNLO--NNLL (short--dashed line), and NNLO--NNNLL (dotted line) results.
}
\label{fig2} 
\end{figure}

\begin{figure}[htb] 
\setlength{\epsfxsize=0.75\textwidth}
\setlength{\epsfysize=0.4\textheight}
\centerline{\epsffile{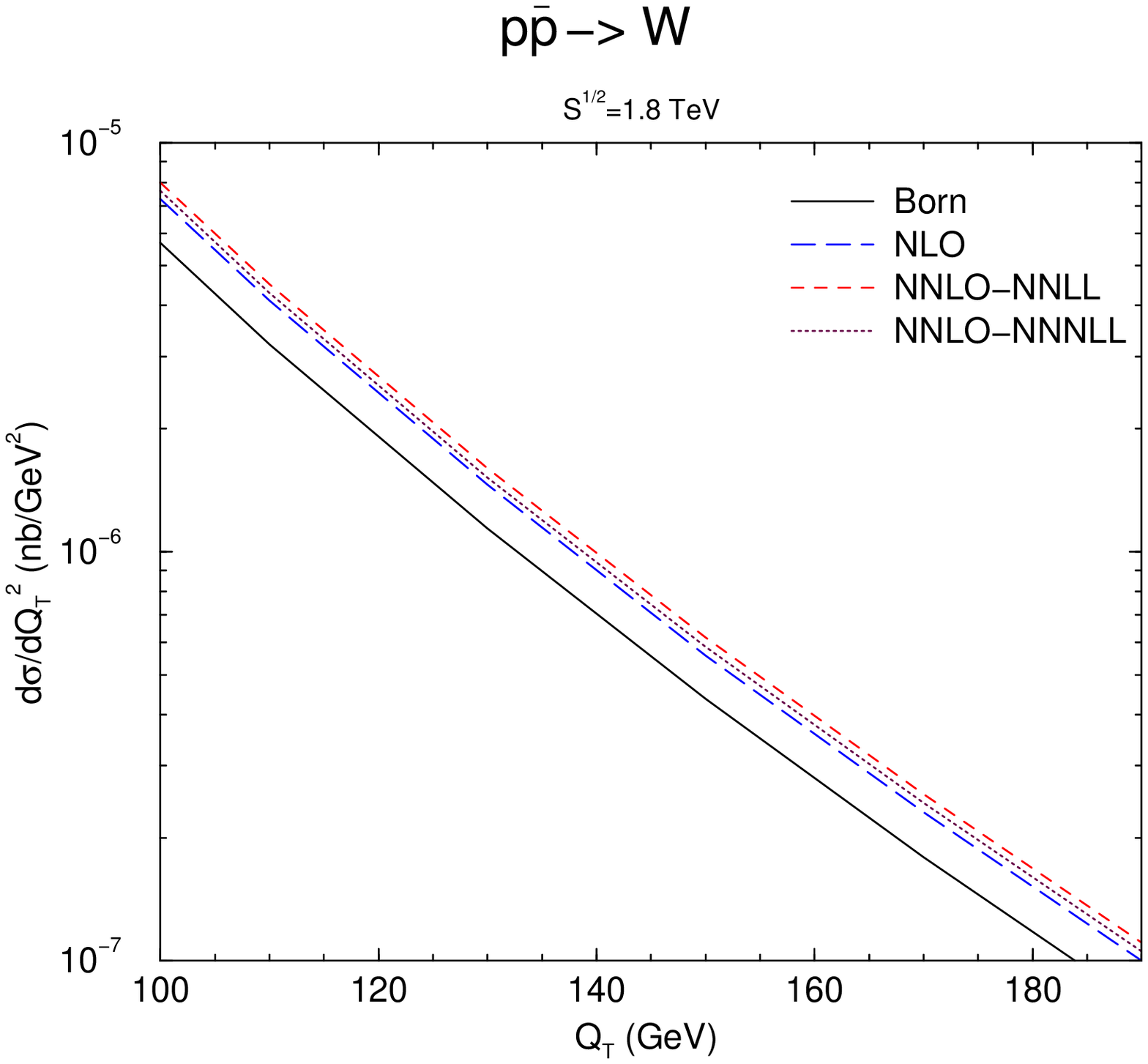}}
\caption[]{The differential cross section,
$d\sigma/dQ_T^2$, of Fig.~1 at high $Q_T$. The labels are
the same as in Fig.~1.}
\label{fig3} 
\end{figure}

\begin{figure}[htb] 
\setlength{\epsfxsize=0.75\textwidth}
\setlength{\epsfysize=0.4\textheight}
\centerline{\epsffile{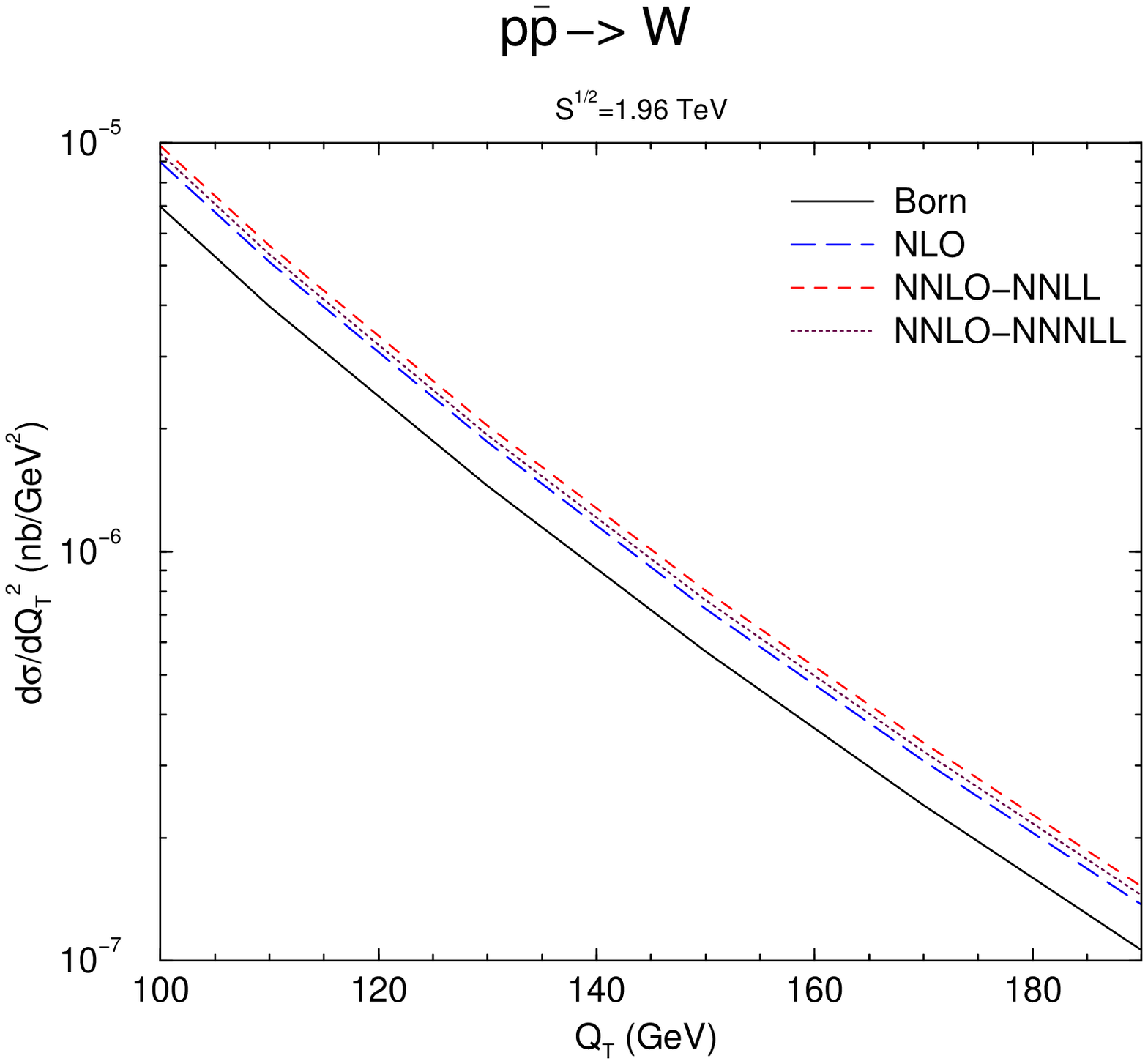}}
\caption[]{The differential cross section,
$d\sigma/dQ_T^2$, of Fig.~2 at high $Q_T$. The labels are
the same as in Fig.~2.
}
\label{fig4} 
\end{figure}

We now apply our results to $W$ hadroproduction at large transverse momentum
at the Tevatron. Throughout we use the MRST2002 approximate NNLO
parton densities \cite{MRST}.
In Fig.~1 we plot the transverse momentum distribution,
$d\sigma/dQ_T^2$, for $W$ hadroproduction at Tevatron Run I 
with $\sqrt{S}=1.8$ TeV. Here we have set $\mu_F=\mu_R=Q_T$.
We plot Born, exact NLO \cite{AR,gpw}, NNLO--NNLL, and 
NNLO--NNNLL results. We see that the 
NLO corrections provide a significant enhancement of the Born
cross section. The NNLO--NNLL corrections provide a further
modest enhancement of the $Q_T$ distribution. 
If we increase the accuracy by including the NNNLL contributions,
which are negative,
then we find that the NNLO--NNNLL cross section lies between
the NLO and NNLO--NNLL results. 
Similar results are shown for $W$ hadroproduction at Tevatron Run II
with $\sqrt{S}=1.96$ TeV in Fig.~2.

Since it is hard to distinguish between the various curves in 
figures 1 and 2, we provide two more figures, Fig.~3
and Fig.~4, for Run I and II respectively, which emphasize
the very high--$Q_T$ region where the soft--gluon approximation holds best.

\begin{figure}[htb] 
\setlength{\epsfxsize=0.75\textwidth}
\setlength{\epsfysize=0.4\textheight}
\centerline{\epsffile{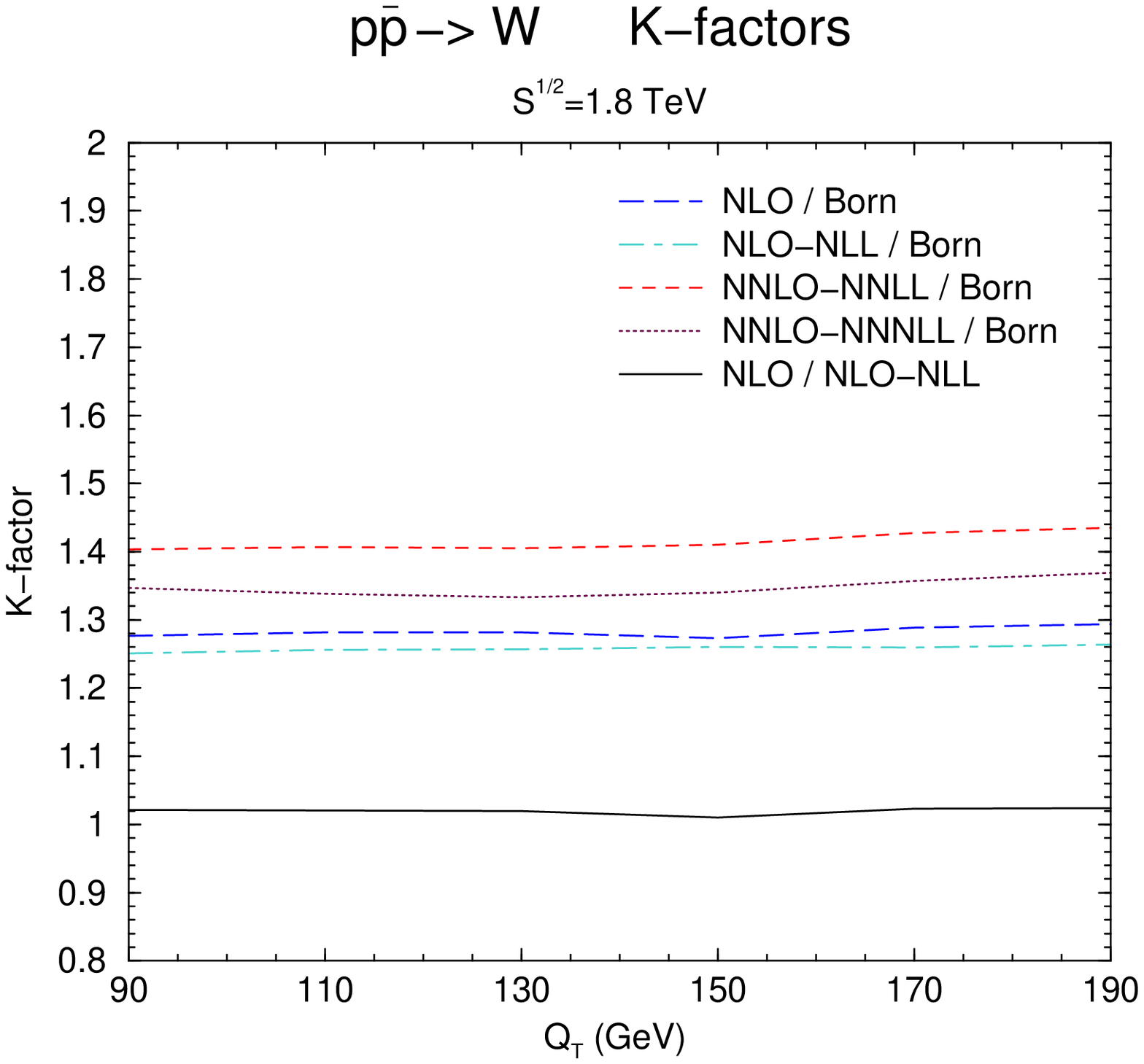}}
\caption[]{The $K$--factors for the differential cross section,
$d\sigma/dQ_T^2$,  for $W$ hadroproduction 
in $p \bar p$ collisions at the Tevatron with $\sqrt{S}=1.8$ TeV 
and $\mu_F=\mu_R=Q_T$.
}
\label{fig5} 
\end{figure}

In Fig.~5 we plot the $K$--factors, i.e. the ratios of
cross sections at various orders and accuracies
to the Born cross section, all with $\mu_F=\mu_R=Q_T$,
in the high--$Q_T$ region of $W$ production at Tevatron Run I.
Shown are the $K$--factors for exact NLO/Born (long--dashed line), 
NLO--NLL/Born (dash--dotted line), NNLO--NNLL/Born (short--dashed line),
and approximate NNLO--NNNLL/Born (dotted line) results.
We also show the ratio of the exact NLO to the NLO--NLL cross section.
It is clear from this line being very close to 1 that the NLO--NLL result
is a very good approximation to the full NLO result, i.e. the
soft--gluon corrections overwhelmingly dominate the NLO cross section. 
The difference between NLO and NLO--NLL is only 2\% for $Q_T > 90$ GeV
and less than 10\% for lower $Q_T$ down to 30 GeV. The fact that the
soft--gluon corrections dominate the NLO cross section
is a major justification for studying the NNLO soft gluon corrections
to this process. We can also see that the various $K$--factors shown in Fig.~5 
are moderate, and nearly constant over the $Q_T$ range shown even though
the distributions themselves span two orders of magnitude in this range.
The NLO corrections are nearly 30\% over the Born, with the NNLO
corrections giving an additional increase, so that the NNLO--NNLL/Born
$K$--factor is around 1.4 and the NNLO--NNNLL/Born $K$--factor is around
1.35.  We note that the $K$--factors for Run II are very similar.

\begin{figure}[htb] 
\setlength{\epsfxsize=0.75\textwidth}
\setlength{\epsfysize=0.4\textheight}
\centerline{\epsffile{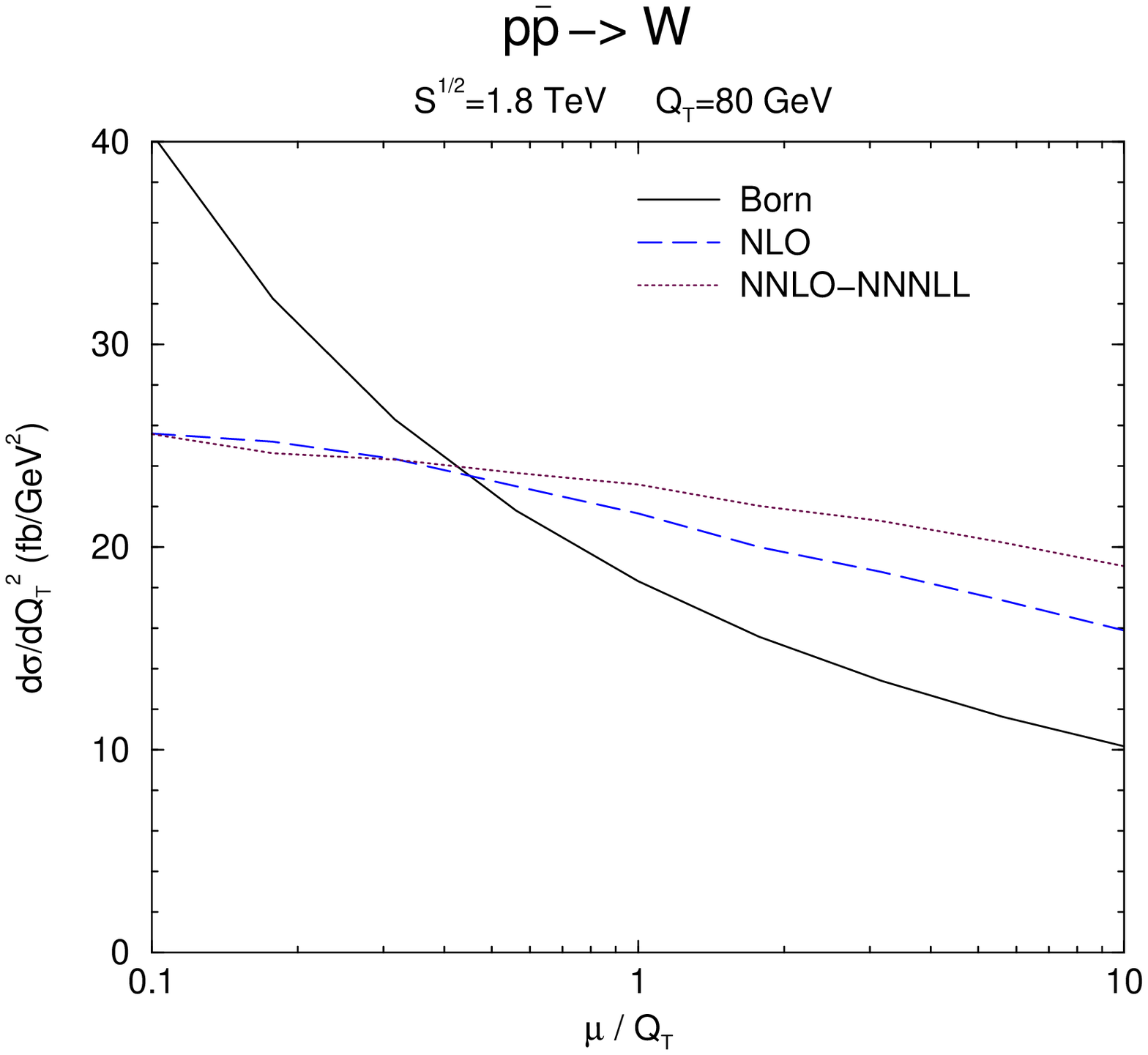}}
\caption[]{The differential cross section,
$d\sigma/dQ_T^2$, for $W$ hadroproduction in $p \bar p$ collisions
at the Tevatron with $\sqrt{S}=1.8$ TeV, $Q_T=80$ GeV, and 
$\mu \equiv \mu_F=\mu_R$.
Shown are the Born (solid line), exact NLO (long--dashed line), 
and NNLO--NNNLL (dotted line) results.
}
\label{fig6} 
\end{figure}

In Fig.~6 we plot the scale dependence of the differential cross section
for $W$ production at Tevatron Run I
for $Q_T=80$ GeV. We define $\mu \equiv \mu_F=\mu_R$ and plot
the differential cross section versus $\mu/Q_T$ over two
orders of magnitude: $0.1 < \mu/Q_T < 10$. We note the good
stabilization of the cross section when the NLO corrections are
included, and the further improvement when the NNLO--NNNLL corrections
(which include all the soft and virtual NNLO scale terms) are added.
The NNLO--NNNLL result approaches the scale independence expected of a truly
physical cross section. 

\begin{figure}[htb] 
\setlength{\epsfxsize=0.75\textwidth}
\setlength{\epsfysize=0.4\textheight}
\centerline{\epsffile{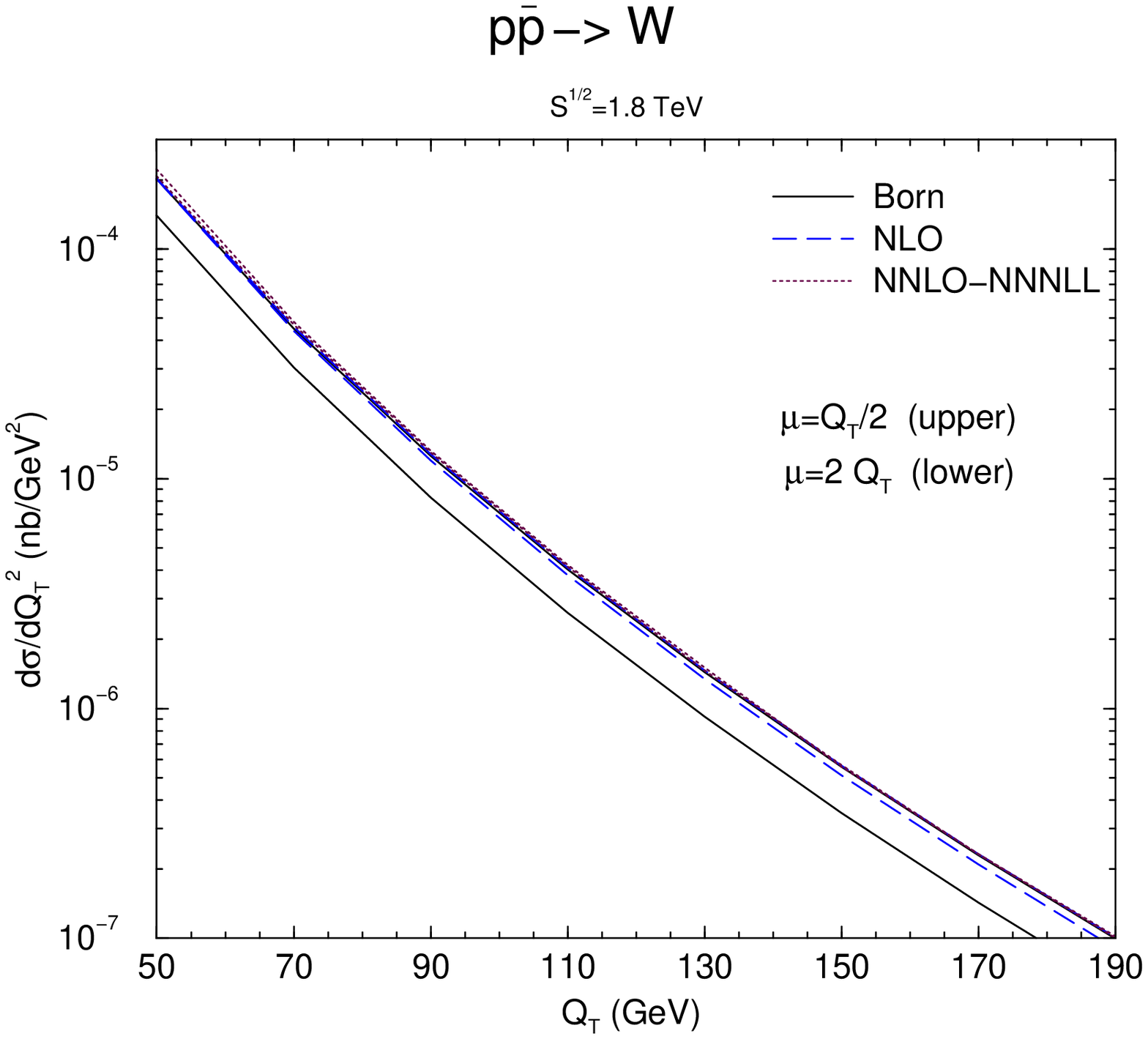}}
\caption[]{The differential cross section,
$d\sigma/dQ_T^2$, for $W$ hadroproduction in $p \bar p$ collisions
at the Tevatron with $\sqrt{S}=1.8$ TeV and 
$\mu\equiv\mu_F=\mu_R=Q_T/2$ or $2Q_T$.
Shown are the Born (solid lines), NLO (long--dashed lines), 
and NNLO--NNNLL (dotted lines) results. The upper lines
are with $\mu=Q_T/2$, the lower lines with $\mu=2 Q_T$.
}
\label{fig7} 
\end{figure}

\begin{figure}[htb] 
\setlength{\epsfxsize=0.75\textwidth}
\setlength{\epsfysize=0.4\textheight}
\centerline{\epsffile{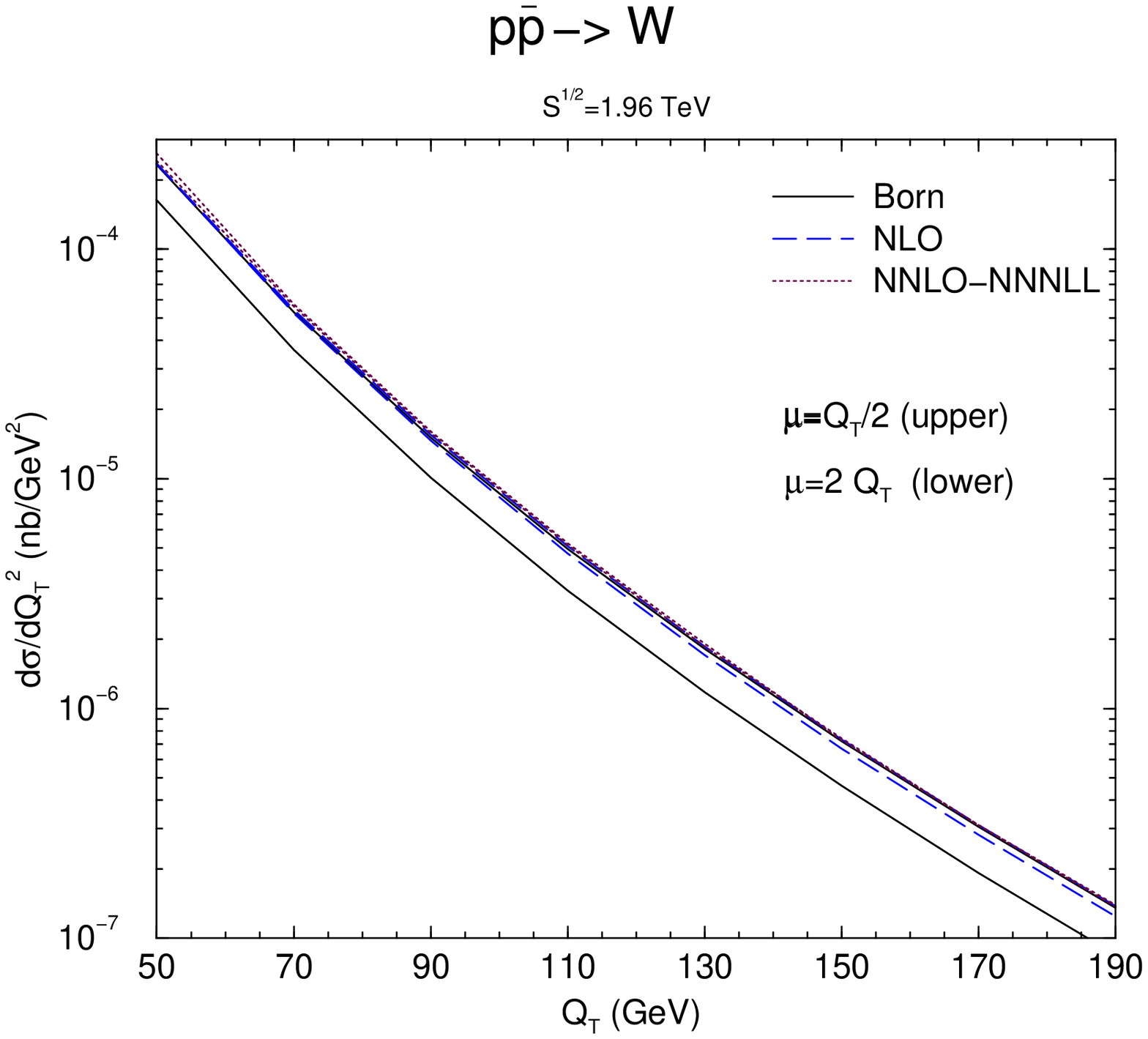}}
\caption[]{The differential cross section,
$d\sigma/dQ_T^2$, for $W$ hadroproduction in $p \bar p$ collisions
at the Tevatron with $\sqrt{S}=1.96$ TeV and 
$\mu \equiv \mu_F=\mu_R=Q_T/2$ or $2Q_T$.
The labels are as in Fig.~7.
}
\label{fig8} 
\end{figure}

In Fig.~7 we plot the differential cross section
$d\sigma/dQ_T^2$ at high $Q_T$ with $\sqrt{S}=1.8$ TeV
for two values of scale, $Q_T/2$ and $2Q_T$,
often used to display the uncertainty due to scale variation.
We note that while the variation of the Born cross section is
significant, the variation at NLO is much smaller, and at
NNLO--NNNLL it is very small. In fact the two NNLO--NNNLL curves
lie on top of the $\mu=Q_T/2$ Born curve, and so does the
NLO $\mu=Q_T/2$ curve. These results are consistent with 
Fig.~6. In Fig.~8 we show analogous results for Run II.

\mysection{CONCLUSION}

The NNLO soft--gluon corrections 
for $W$ hadroproduction at large transverse momentum
in $p{\bar p}$ collisions have been presented, in particular for the case of
 the Tevatron Run I and II.
It has been shown that the NLO soft--gluon corrections fully 
dominate the NLO differential cross section at large transverse momentum,
and that the NNLO soft--gluon corrections provide modest enhancements 
while further decreasing the factorization and renormalization
scale dependence of the transverse momentum distributions as expected in 
any consistent perturbative expansion.

\end{document}